\input epsf.tex
\documentclass[12pt]{article}
\usepackage{graphicx}
\usepackage{amssymb,epsf}

\def\bseq{\begin{subequation}} % = 1a 1b
\def\eseq{\end{subequation}}
\def\bsea{\begin{subeqnarray}} % = 1.1a 1.1b
\def\esea{\end{subeqnarray}}

\newcommand{\beq}{\begin{equation}}
\newcommand{\eeq}{\end{equation}}
\newcommand{\bea}{\begin{eqnarray}}
\newcommand{\eea}{\end{eqnarray}}

\renewcommand{\a}{\alpha}
\renewcommand{\b}{\beta}

\renewcommand{\d}{\delta}
\newcommand{\pa}{\partial}

\newcommand{\g}{\gamma}

\newcommand{\D}{\Delta}

\renewcommand{\l}{\lambda}

\newcommand{\Db}{\bar{D}}

\newcommand{\Phib}{\bar{\Phi}}

\def\Mb{\kern 2pt\mathchoice
            {%displaystyle
             \vbox{\hrule width10pt height 0.4pt depth 0pt
                 \kern 1.2pt\hbox{\kern -2pt$\displaystyle M$}}}
            {%textstyle
                 \vbox{\hrule width10pt height 0.4pt depth 0pt
                 \kern 1.2pt\hbox{\kern -2pt$\textstyle M$}}}
            {%scriptstyle \kern 0.5pt
\vbox{\hrule width6pt height 0.4pt depth 0pt
                 \kern 1.0pt\hbox{\kern -2pt$\scriptstyle M$}}}
            {%scriptscriptstyle \kern 0.5pt
                 \vbox{\hrule width5pt height 0.4pt depth 0pt
                 \kern 0.8pt\hbox{\kern -2pt$\scriptscriptstyle M$}}}}

\def\Sb{\kern 2pt\mathchoice
            {%displaystyle
                 \vbox{\hrule width6pt height 0.4pt depth 0pt
                 \kern 1.2pt\hbox{\kern -2pt$\displaystyle S$}}}
            {%textstyle
                 \vbox{\hrule width6pt height 0.4pt depth 0pt
                 \kern 1.2pt\hbox{\kern -2pt$\textstyle S$}}}
            {%scriptstyle
                 \vbox{\hrule width3.5pt height 0.4pt depth 0pt
                 \kern 1.0pt\hbox{\kern -2pt$\scriptstyle S$}}}
            {%scriptscriptstyle
                 \vbox{\hrule width3pt height 0.4pt depth 0pt
                 \kern 0.8pt\hbox{\kern -2pt$\scriptscriptstyle S$}}}}

\def\Rb{\kern 2pt\mathchoice
            {%displaystyle
                 \vbox{\hrule width5.5pt height 0.4pt depth 0pt
                 \kern 1.2pt\hbox{\kern -2.5pt$\displaystyle R$}}}
            {%textstyle
                 \vbox{\hrule width5.5pt height 0.4pt depth 0pt
                 \kern 1.2pt\hbox{\kern -2.5pt$\textstyle R$}}}
            {%scriptstyle
                 \vbox{\hrule width3.5pt height 0.4pt depth 0pt
                 \kern 1.0pt\hbox{\kern -2.2pt$\scriptstyle R$}}}
            {%scriptscriptstyle
                 \vbox{\hrule width3pt height 0.4pt depth 0pt
                 \kern 0.8pt\hbox{\kern -2.2pt$\scriptscriptstyle R$}}}}

  \def\pp{{\mathchoice
          {%displaystyle
              \kern 1pt%
              \raise 1pt
              \vbox{\hrule width5pt height0.4pt depth0pt
                    \kern -2pt
                    \hbox{\kern 2.3pt
                          \vrule width0.4pt height6pt depth0pt
                          }
                    \kern -2pt
                    \hrule width5pt height0.4pt depth0pt}%
                    \kern 1pt
           }
            {%textstyle
              \kern 1pt%
              \raise 1pt
              \vbox{\hrule width4.3pt height0.4pt depth0pt
                    \kern -1.8pt
                    \hbox{\kern 1.95pt
                          \vrule width0.4pt height5.4pt depth0pt
                          }
                    \kern -1.8pt
                    \hrule width4.3pt height0.4pt depth0pt}%
                    \kern 1pt
            }
            {%scriptstyle
              \kern 0.5pt%
              \raise 1pt
              \vbox{\hrule width4.0pt height0.3pt depth0pt
                    \kern -1.9pt %[e]=0.15pt
                    \hbox{\kern 1.85pt
                          \vrule width0.3pt height5.7pt depth0pt
                          }
                    \kern -1.9pt
                    \hrule width4.0pt height0.3pt depth0pt}%
                    \kern 0.5pt
            }
            {%scriptscriptstyle
              \kern 0.5pt%
              \raise 1pt
              \vbox{\hrule width3.6pt height0.3pt depth0pt
                    \kern -1.5pt
                    \hbox{\kern 1.65pt
                          \vrule width0.3pt height4.5pt depth0pt
                          }
                    \kern -1.5pt
                    \hrule width3.6pt height0.3pt depth0pt}%
                    \kern 0.5pt%}
            }
        }}

  \def\mm{{\mathchoice
                       {%displaystyle
                             \kern 1pt
               \raise 1pt \vbox{\hrule width5pt height0.4pt depth0pt
                                  \kern 2pt
                                  \hrule width5pt height0.4pt depth0pt}
                             \kern 1pt}
                       {%textstyle
                            \kern 1pt
               \raise 1pt \vbox{\hrule width4.3pt height0.4pt depth0pt
                                  \kern 1.8pt
                                  \hrule width4.3pt height0.4pt depth0pt}
                             \kern 1pt}
                       {%scriptstyle
                            \kern 0.5pt
               \raise 1pt
                            \vbox{\hrule width4.0pt height0.3pt depth0pt
                                  \kern 1.9pt
                                  \hrule width4.0pt height0.3pt depth0pt}
                            \kern 1pt}
                       {%scriptscriptstyle
                           \kern 0.5pt
             \raise 1pt \vbox{\hrule width3.6pt height0.3pt depth0pt
                                  \kern 1.5pt
                                  \hrule width3.6pt height0.3pt depth0pt}
                           \kern 0.5pt}
                       }}

\def\pd{{\kern0.5pt
                   + \kern-5.05pt \raise5.8pt\hbox{$\textstyle.$}\kern
0.5pt}}

\def\pmd{{\kern0.5pt
                  \pm \kern-5.05pt \raise6.3pt\hbox{$\textstyle.$}\kern1.5pt}}

\def\md{{\mathchoice
   {%displaystyle
      {{\kern 1pt - \kern-6.2pt \raise5pt\hbox{$\textstyle.$}\kern 1pt}}}
    {%textstyle
      {{\kern 1pt - \kern-6.2pt \raise5pt\hbox{$\textstyle.$}\kern 1pt}}}
    {%scriptstyle
      {\kern0.5pt - \kern-5.05pt \raise3.4pt\hbox{$\textstyle.$}\kern0.5pt}}
    {%scriptscriptstyle
      {\kern0.5pt - \kern-5.05pt \raise3.4pt\hbox{$\textstyle.$}\kern0.5pt}}}}

\newcommand{\ad}{{\dot{\alpha}}}
\newcommand{\bd}{{\dot{\beta}}}
\newcommand{\Del}{\nabla}
\newcommand{\Delb}{\bar{\nabla}}

\newcommand{\newsection}[1]{
\vspace{10mm}
\pagebreak[3]
\refstepcounter{section}
\setcounter{equation}{0}
\setcounter{subsection}{0}
\setcounter{footnote}{0}
\noindent {\bf \thesection. #1}
\nopagebreak
\vspace{2mm}
\nopagebreak}
\renewcommand{\subsection}[1]{
\vspace{5mm}
\pagebreak[3]
\refstepcounter{subsection}
\noindent{ \it \thesubsection. #1} 
\nopagebreak
\vspace{2mm}
\nopagebreak}

\begin{document}

\addtolength{\baselineskip}{.5mm}
\thispagestyle{empty}
\begin{flushright}
October 2003\\
hep-th/0310061\\
CALT-68-2454\\
HUTP-03/A062\\
ITFA-2003-46\\
IFUM-775-FT\\
\end{flushright}

\vspace{8mm}

\centerline{\LARGE Planar Gravitational Corrections}
\bigskip
\centerline{\LARGE For Supersymmetric Gauge Theories}
\vspace{15mm}
\centerline{R. Dijkgraaf,$^1$
M.T. Grisaru,$^2$ H. Ooguri,$^3$ C. Vafa,$^4$ and D. Zanon$^5$}
\vspace{10mm}
\centerline{\it $^1$ Institute for Theoretical Physics \&
Korteweg-de Vries Institute for Mathematics,}
\centerline{\it University of Amsterdam, 1018 XE Amsterdam, The Netherlands}
\medskip
\centerline{\it $^2$ Physics Department, McGill University, }
\centerline{\it Montreal, QC Canada H3A 2T8}
\medskip
\centerline{\it ${}^3$ California Institute for Technology 452-48, Pasadena,
CA 91125, USA}
\medskip
\centerline{\it $^4$
Jefferson Physical Laboratory, Harvard University,}
\centerline{\it Cambridge, MA 02138, USA}
\medskip
\centerline{\it $^5$ Dipartimento di Fisica dell'Universit\'a di Milano,} \centerline{\it INFN, 
Sezione di Milano, Via Celoria 16, I-20133 Milano}

\vspace{10mm}

\centerline{\sc Abstract}

\smallskip

\begin{quote}
\noindent In this paper we discuss the contribution of planar diagrams to
gravitational F-terms for ${\cal N}=1$ supersymmetric gauge theories
admitting large $N$ description. We show how the planar diagrams
lead to a universal contribution at the extremum of the glueball
superpotential, leaving only the genus one contributions, as was
previously conjectured. We also discuss the physical
meaning of gravitational F-terms.
\end{quote}

\newpage

\newsection{Introduction}

It was conjectured in \cite{pert}\ that for ${\cal N}=1$ $U(N)$
theories admitting a large $N$ description, the genus one non-planar
diagrams compute mixed glueball/gravitational F-terms which upon
substitution of the glueball extremum value yield non-perturbative
corrections to gravitational F-terms. This was confirmed in a number
of examples \cite{dijk,mark}.

The genus one contribution to the gravitational coupling was computed
in \cite{gravc} where the idea of $C$-deformation of the chiral ring
\cite{cdeform} played a key role. The method used in
\cite{gravc} involved using  worldsheet techniques as an
inspiration to compute the relevant Feynman diagrams. Similar results
were obtained in \cite{naret,ita}\ using anomaly considerations.

However in addition to the non-planar genus one contribution to gravitational F-terms, it turns out 
that even the planar ones, {\it i.e.} genus zero diagrams, also contribute to gravitational F-terms.
In fact the one-loop planar diagrams in this context were already  computed a long time ago
\cite{onel}.  In this paper we explain how to compute all the planar contributions to gravitational 
corrections. We show that they lead to a universal contribution independent of the coupling 
constants of the theory and consequently they are essentially irrelevant.  We also discuss this 
result from the viewpoint of string theory, in cases where the gauge theory can be obtained on the 
worldvolume of the brane. That the planar contribution can be
 absorbed into a redefinition of glueball fields was also
noted in \cite{naret}.

The organization of this paper is as follows.  In section 2 we present the string inspired 
computation of the gravitational corrections.  In section 3 we check this result for a particular 
example, in the context of more conventional field theory computations. In section 4 we discuss why
the planar contributions sum up to a universal term.  We also discuss the physical interpretation
of gravitational corrections to F-terms.

\newsection{String inspired computation}

As in the papers \cite{cdeform, gravc}, we start our discussion on the string worldsheet. The 
primary goal  is to understand F-terms of ${\cal N}=1$ gauge theories in a
gravitational background. As we will see below, the string theoretical approach simplifies the 
computation since it automatically sums over Feynman diagrams of a given topology. The situation is 
somewhat similar to the one in \cite{bern} where string theory techniques were used to simplify 
gauge theory loop computations. However in our case the relation between gauge theories and string 
theory is more direct.

As shown in \cite{bcov,agnt}, F-terms of a low energy effective theory of
the type II superstring compactified on a Calabi-Yau three-fold $M$ with or without D-branes are 
equal to partition functions of topological string theory. This allows us to compute F-terms of a 
large class of gauge theories obtained as limits of string theory. Moreover, since the string 
theoretical discussion that follows does not deal with details of field contents and their 
interactions -- these are encoded in the choice of the Calabi-Yau space and of the brane 
configurations in it which do not show up explicitly in the discussion below -- the results can
be applied to any ${\cal N}=1$ gauge theory, provided the gauge group is $U(N)$ and all the fields 
are in the adjoint (or fundamental) representations.

The original derivation of \cite{bcov,agnt} was based on the RNS formalism. A more economical 
derivation was given in \cite{bv} using the covariant quantization of the superstring developed in 
\cite{berkcy}. In the formalism of \cite{berkcy} , the four-dimensional part of the worldsheet 
Lagrangian density that is relevant for our discussion is simply given by
\beq
 {\cal L} = {1\over 2} \partial X^\mu \bar \partial X_\mu
  + p_\alpha \bar \partial \theta^\alpha +
p_{\dot \alpha} \bar \partial \theta^{\dot \alpha}+
 \bar{p}_\alpha
\partial \bar \theta^\alpha + \bar p_{\dot \alpha}
\partial \bar \theta^{\dot \alpha},
\label{covariant}
\eeq
where $p$'s are $(1,0)$-forms, $\bar p$'s are $(0,1)$-forms, and
$\theta ,\bar \theta$'s are $0$-forms. The remainder of the
Lagrangian density consists of the topologically twisted ${\cal N}=2$
supersymmetric sigma-model on the Calabi-Yau three-fold and a chiral
boson which is needed to construct the R current. We work in the
chiral representation of supersymmetry in which spacetime
supercharges are given by
\bea
&&Q_\alpha = \oint p_\alpha \nonumber \\
&&Q_{\dot \alpha} = \oint p_{\dot \alpha} - 2i\theta^\alpha
 \partial X_{\alpha \dot \alpha}
 + \cdots,
\label{supercharges}
\eea
where $X_{\alpha\dot \alpha} = \sigma_{\alpha\dot\alpha}^\mu X_\mu$,
and $\cdots$ in the second line represents terms containing
$\theta^{\dot \alpha}$ and $\theta^2 =
\epsilon_{\alpha\beta}\theta^\alpha \theta^\beta$. The second set of
supercharges $\bar Q_\alpha, \bar Q_{\dot \alpha}$ is defined by
replacing $p, \theta$ by $\bar p, \bar\theta$. These generate the
${\cal N}=2$ supersymmetry in the bulk. When the worldsheet is ending
on D-branes and extending into four dimensions, the boundary
conditions for the worldsheet variables are given by
\bea
\label{boundarycondition}
 && (\partial - \bar\partial) X^\mu = 0, \nonumber \\
  && \theta^\alpha = \bar \theta^\alpha, ~~p_\alpha = \bar p_\alpha
\eea
Here we assume that the boundary is located at ${\rm Im}\ z=0$. These
boundary conditions preserve one half of the supersymmetry, generated
by $Q+ \bar Q$.

In these conventions, the vertex operators for the graviphoton $F_{\a
\b}$ and the gravitino $E_{\alpha\beta\gamma}$ field strengths are
given by
\beq
\label{graviphotonvertex}
\int F^{\alpha\beta} p_\alpha \bar p_\beta,
\eeq
and
\beq
\label{gravitinovertex}
\int E^{\alpha\beta\gamma} \left(
 p_\alpha (X \bar \partial X)_{\beta\gamma}
+ \bar p_\alpha (X \partial X)_{\beta\gamma}
+
p_\alpha \bar p_\beta (\theta_\gamma - \bar\theta_\gamma)\right),
\eeq
respectively. Here $(X\partial X)_{\beta\gamma}
= X_{\beta\dot\beta} \partial X_{\gamma\dot\gamma}\epsilon^{\dot\beta
\dot\gamma}$. The gluino ${\cal W}_\alpha$ couples to the boundary
$\gamma_i$ of the worldsheet ($i=1,\cdots,h$) as
\beq
\label{gluinovertex}
\oint_{\gamma_i} {\cal W}^\alpha (p_\alpha
+\bar p_\alpha).
\eeq
Inserting these operators, however, is not the only effect that one has to take into account. It 
was pointed out in \cite{cdeform,gravc} that, in order to preserve the ${\cal N}=1$ supersymmetry, 
one needs modify the chiral algebra of the gluino fields so that they do not anti-commute with each 
other anymore. Rather they have to obey the following $C$-deformed relation,
\beq
\label{cdeform}
\{ {\cal W}_\alpha, {\cal W}_\beta \} =
F_{\alpha\beta} + E_{\alpha\beta\gamma} {\cal W}^\gamma.
\eeq

To discuss the open string theory computation, it is useful to realize
the worldsheet $\Sigma$ of genus $g$ with $h$ boundaries as $\Sigma =
\widetilde{\Sigma}/{\bf Z}_2$, where $\widetilde{\Sigma}$ is a genus
$\tilde{g} = 2g+h-1$ surface without boundary, with ${\bf Z}_2$ acting
as the complex conjugation involution.

\vskip 18pt
\noindent
%---------- FIGURE TOP ------------
\begin{minipage}{\textwidth}
\begin{center}
\includegraphics[width=0.75\textwidth]{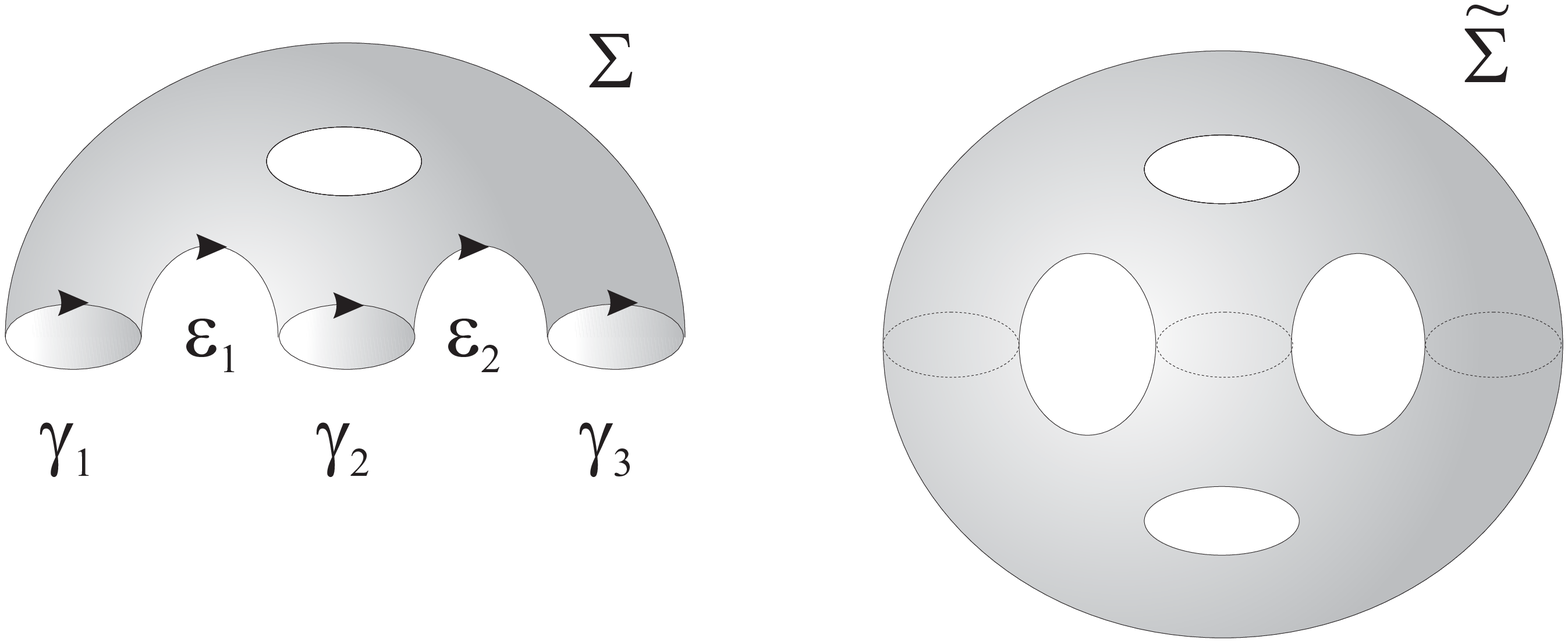}
\end{center}
\begin{center}
{\small{Figure 1: The open string worldsheet $\Sigma$ and its double
$\widetilde\Sigma$}}
\end{center}
\end{minipage}
%---------- FIGURE END ------------

\vskip 20pt

\noindent The boundaries of $\Sigma$ are fixed point sets of the ${\bf Z}_2$
involution. Let us choose a basis of homology cycles of
$\tilde{\Sigma}$ as $\{ A_a, B_a, ~(a=1,\ldots,g),~
\gamma_i, \epsilon_i ~(i=1,\ldots,h-1) \}$, where $\gamma_i$ is
the cycle around the $i$-th boundary, $\epsilon_i$ is an interval
connecting $\gamma_i$ and $\gamma_{i+1}$, so that their intersections
are $(A_a, B_b) = \delta_{ab}$, $(\gamma_i, \epsilon_j) = \delta_{ij}-
\delta_{h,j}$ and otherwise $=0$.

Without other operators in the bulk, we can use the Riemann bilinear identity as in \cite{cdeform}\ 
to rewrite the surface integral of the graviphoton vertex operator as
\bea
 \int_\Sigma F^{\alpha\beta} p_\alpha \bar p_\beta
&=& \sum_{a=1}^g 2F^{\alpha\beta} \oint_{A_a} (p_\alpha+\bar p_\alpha)
\oint_{B_a} (p_\beta +\bar p_\beta) \nonumber \\
&+&\sum_{i=1}^{h-1} 2F^{\alpha\beta} \oint_{\gamma_i} (p_\alpha+\bar p_\alpha)
\int_{\epsilon_i} (p_\beta +\bar p_\beta).
\eea
Similarly the surface integral of the gravitino vertex operator can be
re-expressed as
\bea
\label{gravriemann}
&&
 \int_\Sigma E^{\alpha\beta\gamma} \left(
 p_\alpha (X\bar \partial X)_{\beta\gamma}
+ \bar p_\alpha (X \partial X)_{\beta\gamma}
+ p_\alpha \bar p_\beta (\theta_\gamma - \bar\theta_\gamma)\right)
\nonumber \\
&&= \sum_{a=1}^g
 E^{\alpha\beta\gamma}
\oint_{A_a} (p_\alpha + \bar p_\alpha)
\oint_{B_a} \left( (X\partial X)_{\beta\gamma} -
(X\bar\partial X)_{\beta\gamma}
+ p_\beta \theta_\gamma - \bar p_\beta \bar \theta_\gamma\right)
\nonumber \\
&&~~~+\sum_{a=1}^g
 E^{\alpha\beta\gamma}
\oint_{B_a} (p_\alpha + \bar p_\alpha)
\oint_{A_a} \left( (X\partial X)_{\beta\gamma} -
(X\bar\partial X)_{\beta\gamma}
+ p_\beta \theta_\gamma - \bar p_\beta \bar \theta_\gamma\right)
\nonumber \\
&&~~~+ \sum_{i=1}^{h-1}
E^{\alpha\beta\gamma}
\oint_{\gamma_i} (p_\alpha + \bar p_\alpha)
\int_{\epsilon_i} \left( (X\partial X)_{\beta\gamma} -
(X\bar\partial X)_{\beta\gamma}
+ p_\beta \theta_\gamma - \bar p_\beta \bar \theta_\gamma\right) .
\eea
Note that we do not have terms coming from exchanging $\gamma_i$ and $\epsilon_i$ in the last line 
since $\partial X = \bar \partial X$ and $p=\bar p,~ \theta=\bar\theta$ on the boundaries. These 
terms remain if we have the gluino vertex operator (\ref{gluinovertex}) on the boundary since it 
has non-zero correlations with $\theta$ and $\bar \theta$ in the gravitino vertex operator. However 
this effect is cancelled if we turn on the $C$-deformation (\ref{cdeform}).

We are interested in terms of the form $E^2 S^{h-2}$ with $S={\rm
Tr}\, {\cal W}_\alpha {\cal W}^\alpha$. Let us analyze planar
diagrams (so $g=0$) with $h$ boundaries. On a planar diagram there
are no $A_i,B_i$ cycles and so the integral of the gravitino vertex is
given by
\beq
\label{final}
\sum_{i=1}^{h-1}
E^{\alpha\beta\gamma}
\oint_{\gamma_i} (p_\alpha + \bar p_\alpha)
\int_{\epsilon_i} \left( (X\partial X)_{\beta\gamma} -
(X\bar\partial X)_{\beta\gamma}
+ p_\beta \theta_\gamma - \bar p_\beta \bar \theta_\gamma\right) \eeq
Note that we have the same factor $\oint_{\gamma_i} (p_\alpha +\bar
p_\alpha)$ as in the gluino vertex (\ref{gluinovertex}) . The
difference is that, whereas the gluino vertex operators carries an
additional group theoretical factor, the gravitino vertex includes the
integral along the interval $\epsilon_i$,
\beq
\label{lorentz}
M_{\beta\gamma}=\int_{\epsilon_i} \left( (X\partial X)_{\beta\gamma} -
(X\bar\partial X)_{\beta\gamma}
+ p_\beta \theta_\gamma - \bar p_\beta \bar \theta_\gamma\right).
\eeq
It acts as a generator of Lorentz transformations on the open string connecting the two boundaries 
$\gamma_i$ and $\gamma_{i+1}$.  (The Lorentz generator $M_{\mu\nu}$, which is antisymmetric in 
$\mu,\nu=0,\cdots,3$, can be decomposed as $M_{\alpha\dot\alpha \beta\dot\beta} = 
M_{\alpha\beta}\epsilon_{\dot\alpha\dot\beta} + M_{\dot\alpha\dot\beta} \epsilon_{\alpha\beta}.$ We 
can identify $M_{\alpha\beta}$ in this decomposition as the operator (\ref{lorentz}).)

At this stage, it is useful to compare the computation of the $E^2
S^{h-2}$ term with the $S^{h-1}$ term coming from the same surface
with $h$ boundaries \cite{pert, DGLVZ}: The latter is the standard
superpotential term in the ${\cal N}=1$ gauge theory, while the former
is its gravitational correction. In both cases, the $(h-1)$ zero modes
of $p_\alpha$ are absorbed by $\oint_{\gamma_i} p_\alpha$
$(i=1,\cdots,h-1)$ in the vertex operators. For the $S^{h-1}$ term,
after absorbing the fermion zero modes, we are left with taking traces
over gauge group indices around $h$ boundaries; $(h-1)$ of them give a
factor $S$ and the remaining one gives a factor of $N$, the rank of
the gauge group. In addition, there is a combinatorial factor of $h$
due to the choice of one out of the $h$ boundaries where we do not
insert the gluino. 

For the $E^2 S^{h-2}$, there will be $(h-2)$ boundaries on which we
have two gluino insertions each. In addition, we have two insertions
of the Lorentz generator $M_{\alpha\beta}$ defined by (\ref{lorentz})
. The operator product singularities between the $X$'s in the two
$M_{\alpha\beta}$ are cancelled by those between $p$ and $\theta$.
(This of course should have been the case due to the topological
nature of the worldsheet theory.) Moreover, the zero modes of $p$
have already been absorbed. So, the computation reduces to an integral
over the momentum zero modes of $X$. Due to its topological nature,
the computation is essentially the same as the one for the one loop case as in
\cite{VafaFJ}, and produces the contraction of $E_{\alpha\beta\gamma}
E^{\alpha\beta\gamma}$. In addition, there is a factor of $N^2$ coming
from the gauge group trace over the two boundaries and $h(h-1)$ due to
the choice of these two boundaries. Therefore, while the standard
superpotential for $S$ takes the form $N
\partial {\cal F}_0/ \partial S$, the gravitational correction takes the
form 
$$
{\cal L} = E_{\alpha\beta\gamma} E^{\alpha\beta\gamma} N^2 {\partial^2
{\cal F}_0 \over \partial S^2}.
$$
More generally, if we have various different boundary types where the gauge group is broken as
$$
U(N) \to U(N_1) \times \cdots \times U(N_k),
$$
the same reasoning as above shows that we obtain for the gravitational
correction
\beq
{\cal L} = E_{\alpha\beta\gamma} E^{\alpha\beta\gamma} \sum_{i,j}
N_i N_j {\partial^2 {\cal F}_0
\over \partial S_j \partial S_j}.
\label{planar}
\eeq

It is useful to compare the string theoretical computation here to
a field theoretical computation.  Needless to say all the steps here
could be given a field theoretic flavor simply by
considering the $\alpha'\rightarrow 0$ version of the same arguments,
though it would be cumbersome.  The especially non-trivial fact is the
use of the Riemann bilinear identity in organizing the sum of various
field theory diagrams into a simple expression.  At any rate it would
be useful to check, at least in some examples, these results with those
of more conventional field theoretic techniques.  This we will do for
a non-trivial two loop computation in the next section.  As we
will see, the field theoretical computation has two ingredients:
one is an effect due to explicit insertions of gravity vertices
(Figure 3) and another is due to the
gravitational $C$-deformation
$\{ {\cal W}_\alpha, {\cal W}_\beta \} = E_{\alpha\beta\gamma} {\cal
W}^\gamma$ in diagrams involving only insertions of the Yang-Mills
fields (Figure 2c).

Let us explain how these ingredients also arise in the string theory
computation. One starts with the vertex operator for the gravitino
(\ref{gravitinovertex}), but one has to combine it with the
$C$-deformation, which is necessary for preservation of supersymmetry
due to the fermionic part of the vertex \cite{cdeform}, in order to be able to
write its surface integral as the sum of contour integrals as in
(\ref{gravriemann}). This is then used, in the case of $g=0$, to
arrive at the expression given above. Note that it is important that
the gravitino vertex (\ref{gravitinovertex}) has two types of terms,
one of the form $p X \partial X$ and another of the form $p\bar p
(\theta-\bar\theta)$. It is the cancellation of the effects from these
two types  that maintains the topological BRST invariance on
the worldsheet. For example, there is no operator product singularity
between two gravitino operators of this type since the singularity
coming from contractions of $X$'s is cancelled by that coming from
contractions of $p$ and $\theta$. One can also see, in the field
theory limit, that the $p X \partial X$ term contributes in the
Feynman diagrams involving explicit insertions of gravitino vertices
as in Figure 3, while the $p\bar p (\theta-\bar\theta)$ term
gives rise to the $C$-deformation and therefore is responsible for
diagrams such as Figure 2c. In the string theory computation, these
two effects are combined into the single expression (\ref{final}),
from which we can read off the final result directly.

\newsection{Planar Two-loop Calculation}

In ref. \cite{DGLVZ}\ the effective glueball superpotential was
computed in a perturbative field theory calculation, which led to a
justification of the conjecture in \cite{pert}. The model considered
there, which incorporates all the relevant features, consists of a
chiral matter superfield in the adjoint representation with action
\beq 
S= {\rm Tr}\left[\int d^4x d^4
\theta ~\Phib \Phi + \int d^4x d^2 \theta \left(
\frac {m}{2} \Phi^2 + \frac{\l}{3!} \Phi^3\right) + {\rm
h.c.}\right] \label{action} 
\eeq 
The coupling to the background Yang-Mills (YM) superfield was achieved
by requiring $\Phi$ to be {\it covariantly} chiral, {\it i.e}
$\Delb_\ad \Phi=0$, where $\Delb_\ad = \Db_\ad - i \bar{\Gamma}_\ad$
is a derivative covariantized through the superspace YM connection
$\bar{\Gamma}_\ad$.

It was shown there that order by order in perturbation theory one can integrate out the matter 
fields and compute the corresponding contribution to the superpotential of the gluino condensate.
It is obtained from planar graphs and takes the form, at $L$ loops, $({\rm Tr}\, {\cal W}^2)^L$ 
where the Yang-Mills superspace field strength ${\cal W}_\a$  is evaluated at zero momentum.

Here we want to show that similar techniques can be employed when the matter fields are coupled to 
a supergravity background as well.  We do this explicitly at two loops for planar index graphs and 
show that the perturbative field theory computation reproduces what is expected from the string 
theory approach of the previous section.
Namely, we obtain a result proportional to ${\cal W}^2E^2$ where $E^2 =
\frac{1}{2} E^{\a \b \g}E_{\a \b \g}$. (It turns out that in order to obtain this result
in a standard field theory calculation it is crucial to consider a {\em nonabelian} YM background.  
This is done in order to implement the $C$-deformation of the chiral ring \cite{cdeform, gravc}\ in the
context of conventional field theory computations.)  There are two rather distinct sources for the 
${\cal W}^2E^2$ contributions and we consider them in turn.

The presence of supergravity adds new features to the calculations of ref. \cite{DGLVZ} where it 
was argued that only the first two diagrams of Fig. 2, drawn in 't Hooft double line notation with 
dots indicating insertions of ${\cal W}_\a$ factors, contribute to the $\int d^2 \theta ({\rm 
Tr}\,{\cal W}^2)^2$ superpotential.  For such a contribution it sufficed to consider objects in the 
chiral ring, {\it i.e.} objects which are annihilated by the $\Delb_\ad$ spinor derivative modulo 
local and gauge invariant $\Delb$-exact terms, which would not contribute to the chiral integral.
In the absence of supergravity, because of the chiral ring relation \beq \{{\cal W}_\a, {\cal
W}_\b\} =0 ~~{\rm mod} ~~\Delb \label{chiralring} \eeq which follows from $\{\Delb^\ad ,[\Del_{\a 
\ad}, {\cal W}_\b]\}=-2\{{\cal W}_\a, {\cal W}_\b\}$, the only relevant object was the gluino
condensate ${\rm Tr}\,{\cal W}^2$ while higher traces (in particular ${\rm Tr}\,{\cal W}^4 $) 
vanish. This implied that not more than one pair of ${\cal W}_\a$ could be inserted in a given 
index loop.  However, with supergravity present this is no longer the case \cite{gravc}, as we will 
now review.

\vskip 18pt
\noindent
%---------- FIGURE TOP ------------
\begin{minipage}{\textwidth}
\begin{center}
\includegraphics[width=0.75\textwidth]{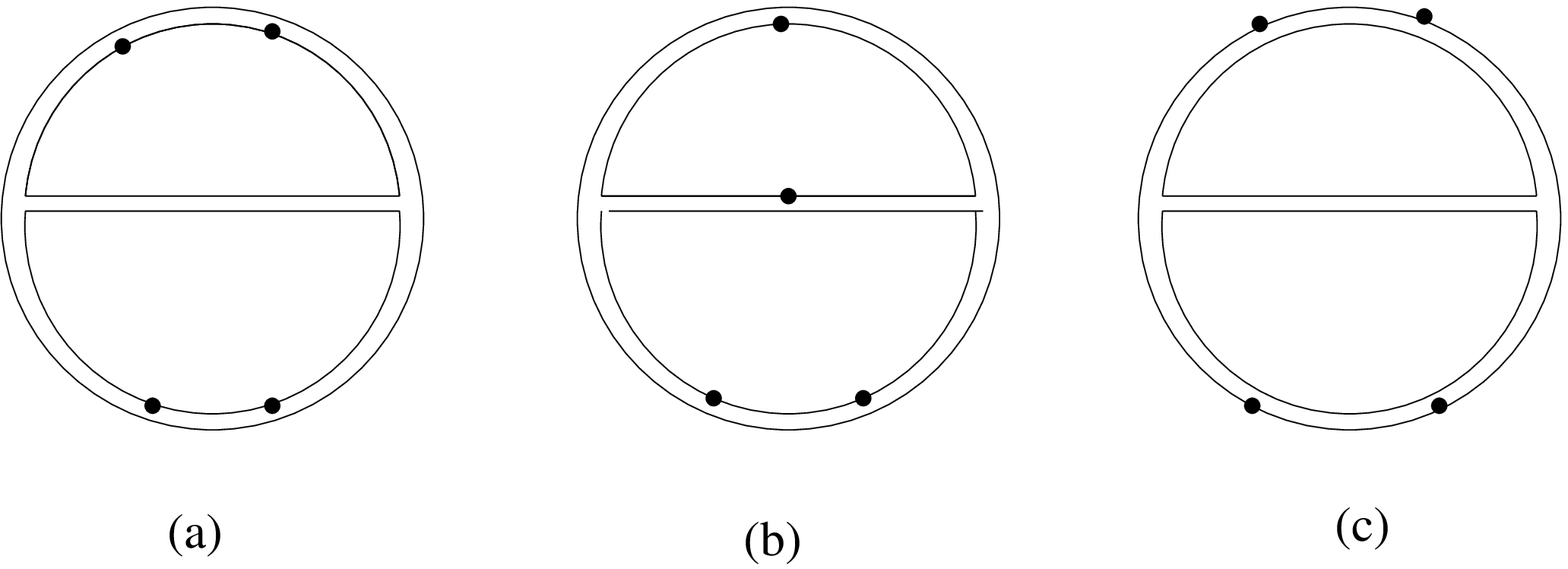}
\end{center}
\begin{center}
{\small{Figure 2: Two-loop diagrams for SSYM}}
\end{center}
\end{minipage}
%---------- FIGURE END ------------

\vskip 20pt

{}From the algebra of superspace covariant derivatives satisfying the usual constraints, and as a 
consequence of the Bianchi identities one finds \beq [\Delb_\ad , \Del_{\b \bd}] = C_{\ad \bd} 
{\cal W}_\b +C_{\ad \bd} {E_{\b \g}}^\d {M_\d}^\g, \eeq where ${M_\d}^\g$ is the Lorentz generator
we have introduced in the previous section, with $[M_{\a \b}, \psi_\g] = \frac{1}{2} C_{\g \a} 
\psi_\b +\frac{1}{2} C_{\g \b} \psi_\a$ and $C_{\a\b}= i \epsilon_{\a\b}$ ( $C^{\a \b}C_{\a\b} = 2$
and similarly for dotted indices). One derives then, using the Jacobi identity and the
chirality of ${\cal W}_\a$  \bea
\{\Delb^\ad ,[\Del_{\a \ad}, {\cal W}_\b]\} &=& \{[\Delb^\ad, \Del_{\a \ad}],{\cal W}_\b\} \\
&=& -2\{{\cal W}_\a, {\cal W}_\b\} -2\{{E_{\a \g}}^\d {M_\d}^\g, {\cal W}_\b \}\nonumber\\
&=& -2\{{\cal W}_\a, {\cal W}_\b\}-2{E_{\a \b}}^\g {\cal W}_\g. \nonumber \eea Therefore, in the 
presence of background supergravity, one has \beq \{{\cal W}_\a, {\cal W}_\b\}={E_{\a \b\g}} {\cal 
W}^\g ~~{\rm mod} ~~\Delb. \label{chiralringgrav} \eeq We emphasize that in order to obtain a 
modified chiral ring relation as in (\ref{chiralringgrav}) it is crucial to consider a {\em 
nonabelian} YM background. In this case no special deformation is required.  Otherwise the 
implementation of the $C$-deformation would have to be realized in an unconventional fashion in field 
theory diagrams by introducing suitable boundary terms as in \cite{cdeform}.

It is easy to show from (\ref{chiralringgrav})  that \beq {\cal W}^\a {\cal W}_\a
{\cal W}^\b {\cal W}_\b = -\frac{1}{12} E^{\a \b \g}E_{\a \b \g} {\cal W}^\rho {\cal W}_\rho \eeq
In particular, if the computation of a Feynman amplitude produces the trace of the left-hand-side
above, we can replace the result by the corresponding trace of the right-hand-side. Indeed, Fig.2c
with four ${\cal W}$'s inserted in the same index loop gives rise to such a contribution.  
Therefore, at the two-loop level, in addition to the previously computed $({\rm Tr}\,{\cal 
W}^2)^2$, we will obtain a term ${\rm Tr}\,{\cal W}^4 \sim E^2\,{\rm Tr}\,{\cal W}^2$

We consider therefore the diagram  in Fig.2c.  As in \cite{DGLVZ} the calculation is
performed using a Schwinger parametrization for the propagators \beq \langle \Phi \Phi \rangle_i = 
\int_0^\infty \!\! ds_i \ \exp\left[-s_i\left(p_i^2+{\cal W}^\a\pi_{i\a} + m\right)\right]. \eeq 
Here we have set to zero the explicit supergravity dependence since it does not enter this part of 
the calculation; the coupling to supergravity is through the covariant derivatives in terms of 
which the superfield strength ${\cal W}_\a$ is defined. Also we went to Fourier transforms with 
respect to both space-time and spinor derivatives, introducing thus the corresponding momentum 
operators $p$ and $\pi$. Finally, we have set $\bar{m}=1$ since, by holomorphy arguments, one knows 
that it does not enter the final result. The actual supergraph manipulation is essentially the same 
as for Fig.2a and can be found in \cite{DGLVZ}.  With a labeling $(s,t,u)$ for the three Schwinger
parameters, after taking into account factors for combinatorics, (1/2), and group theory, ($2N^2$), a
factor of $(1/2)^2$ from the second order expansion of exponentials and a $1/3$ from a 
symmetrization over the Schwinger parameters, we find a net contribution \bea && \frac{N^2}{12}  
{\rm Tr}({\cal W}^\a {\cal W}_\a {\cal W}^\b {\cal W}_\b)
\frac{s^2t^2+t^2u^2+u^2s^2}{\Delta^2}\nonumber\\
&& =-\frac{N^2}{144}E^{\a \b \g} E_{\a \b \g} {\rm Tr}({\cal W}^\d {\cal W}_\d) \frac{s^2t^2+{\rm 
permutations}}{\Delta^2} \label{WWWW} \eea where \beq \Delta=st+su+ut \eeq This is to be multiplied 
by $e^{-m(s+t+u)}$ and integrated over the Schwinger parameters.

 We turn now to the calculation of
supergraphs with explicit insertions of supergravity vertices.
As in cite{DGLVZ} we make use of a background covariant formulation
of the theory, extended to the case of background supergravity
\cite{GZ1,GZ2}. This allows us to perform the Feynman diagram
computation using covariant supergraph rules which simplify the
algebra in a drastic manner. We start again with the action in
(\ref{action}) where now the chiral superfield is covariantly chiral
with respect to both Yang-Mills and supergravity, {\it i.e.}\ the
spinor derivatives $\Del_\a$ and $\Delb_\ad $ are covariantized with
respect to both. We assume that the background is on shell. As in
\cite{DGLVZ} corrections to the superpotential are obtained by
computing $\int d^2\theta$ terms from vacuum diagrams with quantum
vertices
\beq 
\frac{\l}{3!}\int d^4x d^2 \theta ~ \Phi^3 \label{vertex}
\eeq
from the action in (\ref{action}) and propagators
\beq
\langle \Phi\Phi \rangle =-\frac{\bar{m}}{\Box_+ -m \bar{m}}
 \label{propagator}
\eeq The dependence on the  external fields is  contained in \beq \Box_+=\frac{1}{2}\Del^a \Del_a- 
i{\cal W}^\a \Del_\a \label{covbox} \eeq where $\Del_a=-i\{\Del_\a,\Del_\ad\}= {E_a}^M D_M + {\rm
connections}$, $M=\{m, \mu , \dot{\mu}\}$, and ${E_A}^M$ is the supergravity vielbein.

In \cite{GZ2} we have shown that $\frac{1}{2}\Del^a\Del_a= \frac{1}{2} E^{aM}D_M {E_a}^N D_N$ can be
expanded with respect to spinor derivatives so as to take the form
\beq 
\frac{1}{2}\Del^a\Del_a =
\frac{1}{2}D^a D_a -A^\a \Del_\a -\bar{A}^\ad \Delb_\ad -B\Del^2 
-\bar{B}\Delb^2 -C^{\a
\ad}[\Del_\a , \Delb_\ad ] \label{spingrav}
\eeq
where, with the supergravity fields on shell,
\bea
&&A^\g = e^{a\g}D_a -(D^a e_a^{~\g})
~~~,~~\bar{A}^{\dot{\g}} =e^{a \dot{\g}} D_a - (D^a e_a ^{~\dot{\g}})
\nonumber\\
&&B= \frac{1}{2}e^{a\g}e_{a \g} ~~,~~ \bar{B}=\frac{1}{2} e^{a \dot{\g}}e_{a \dot{\g}} ~~,~~ C^{\g
\dot{\g}}= \frac{1}{2}e^{a \g}e_a^{~\dot{\g}} \label{supergravity}
\eea
(However, all the terms containing $\Delb_{\ad}$ do not contribute
here and will be dropped henceforth.) As explained in \cite{GZ2},
$D_a$ is a space-time ``covariant'' derivative. At $\theta=0$ and in
Wess-Zumino gauge it reduces to the ordinary gravitational covariant
derivative. The superfield $e^{a \g}$ is the basic object we work
with (not to be confused with the vector-spinor part of the original
vielbein $E^{aM}$, although the two are equal at the linearized level); its
first component is the gravitino field.

When computing vacuum diagrams with vertices from (\ref{vertex}) and propagators in 
(\ref{propagator}), the background dependence is obtained by expanding the propagators. In this way 
one produces factors of $\Del_\a$ which are needed to complete the covariant $D$-algebra at every 
loop through the rule \beq \d^{(2)}(\theta-\theta')\Del^2 \d^{(2)}(\theta-\theta')=1 
\label{Dalgebra} \eeq The external YM fields are contained in the explicit superfield strength 
${\cal W}_\a$, while the relevant supergravity fields appear through terms in 
(\ref{supergravity}).  Although these vertices are not in covariant form, the invariance of the 
action under general coordinate and local supersymmetry transformations (at the linearized level, 
we have invariance under the gauge transformations $\d {e_a}^\g= \pa_a K^\g$ \cite{GZ2}) guarantees 
that the final result of our calculation will be expressible (on shell) in terms of the field 
strength $D_{a}{e_b}^\g - D_{b}{e_a}^\g$. We note here the relation \beq D_{[a}e_{b]\g} = iC_{\bd 
\ad} E_{\a\b\g} + iC_{\b \a} E_{\ad\bd\dot{\g}} \label{gravcov} \eeq

The noncovariance of the supergravity vertices makes the supergraph calculation rather 
complicated.  In particular, unlike the YM case where we could from the very beginning set the
momenta of the external fields ${\cal W}_{\a}$ to zero, here, since the couplings to supergravity 
are proportional to the ``potential'' ${e_a}^\g$ rather than the field strength $D_{[b}{e_{a]}}^\g$ 
we cannot set immediately the gravitational external field to zero momentum.  A brief description 
of the steps required is as follows:

a) As in the SSYM
case we can carry out the rather trivial D-algebra on the supergraphs,
but the complications arise from the presence of momentum factors in
the numerator of the resulting Feynman integrals.

b) Using gauge invariance we project out, on each diagram, a part
which is sufficient for reconstructing the full result which involves
now, in the numerator, scalar products of the loop momenta. At this
stage we can set the external momenta to zero.

c) After writing the propagators in exponential, Schwinger parameter,
form we replace these scalar products with derivatives with respect to
the parameters, after which the momentum integrals are easily carried
out leaving us with a standard expression $\Delta(s_i)^{-2}e^{-m\sum
s_i}$ multiplied by some additional dependence on the $s_i$.

We give
now some details. At the two-loop level we are working with supergraphs with a total of four spinor $\Del_\a$
derivatives obtained from the expansion of the three propagators using 
(\ref{propagator},\ref{covbox},\ref{spingrav}), with two Yang-Mills field strengths $ {\cal W}^\a 
\Del_\a $ and one or two insertions of supergravity fields $B\Del^2$ and $A^\a \Del_\a$ 
respectively.  The various supergraphs are described by the diagrams in Fig.2 where the dots 
indicate Yang-Mills insertions and the lines supergravity insertions. These insertions produce the 
necessary number of spinor derivatives for carrying out the trivial D-algebra.

\vskip 18pt
\noindent
%---------- FIGURE TOP ------------
\begin{minipage}{\textwidth}
\begin{center}
\includegraphics[width=0.75\textwidth]{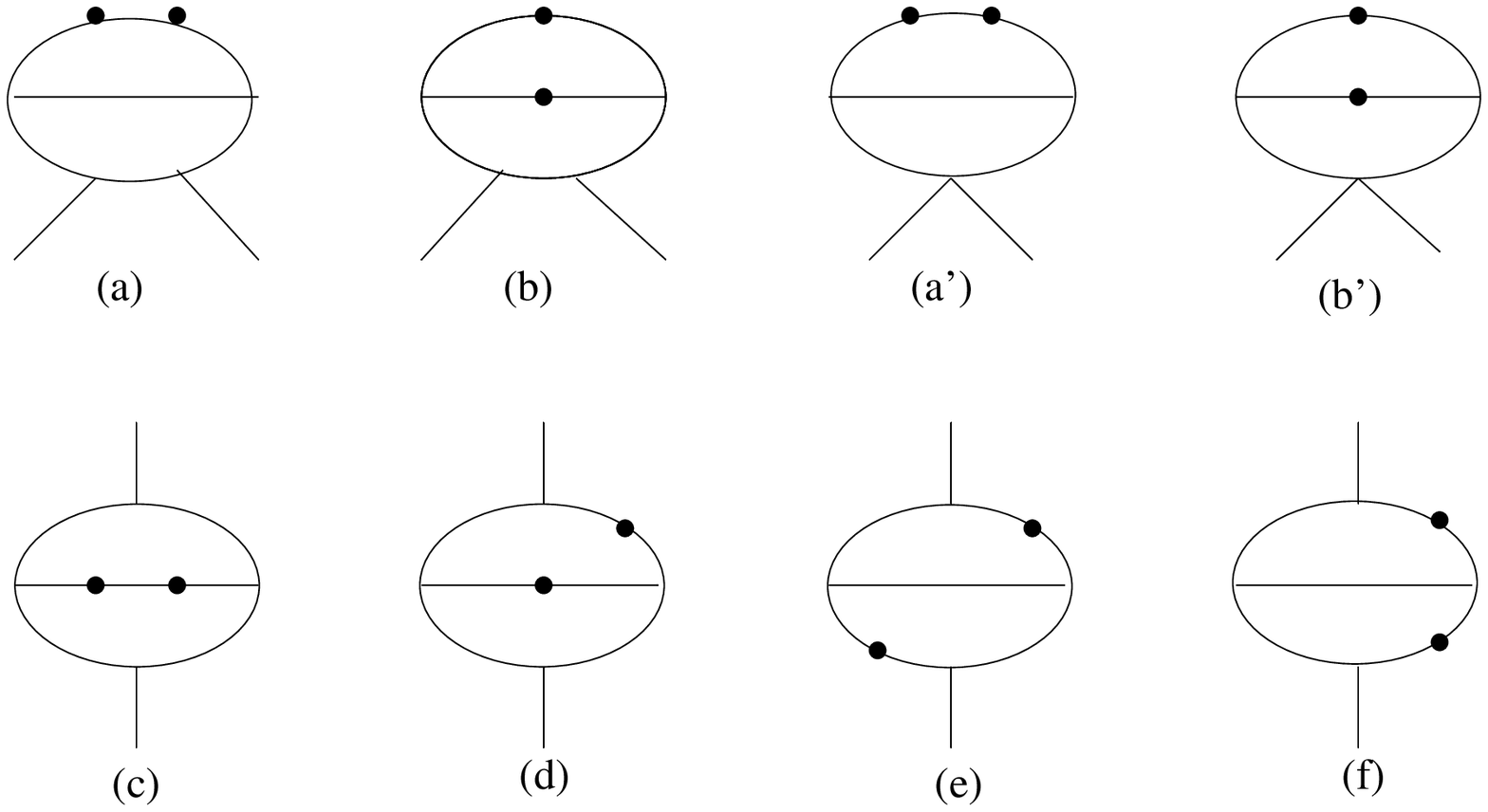}
\end{center}
\begin{center}
{\small{Figure 3: Two-loop diagrams for the supergravity-SSYM system}}
\end{center}
\end{minipage}
%---------- FIGURE END ------------

\vskip 20pt

The YM fields are already in covariant form while the supergravity
fields are not. Therefore when computing the various Feynman diagrams,
we have to do the momentum integration with insertions at zero
momentum for the Yang-Mills fields, but at nonzero momentum $q$ for
the supergravity fields. However, in momentum space, the final
supergravity covariant result must take the form
\bea
 \Gamma &=& e^{a\g}(-q)( q^2 \d_{ab} -
q_aq_b) {e^b}_\g(q)G(q^2) =\frac{1}{2}(e^{a\g} q^b- e^{b \g}q^a)(e_{a\g}q_b-e_{b\g}q_a)G(q^2)
 \nonumber\\
&=&\frac{1}{2} C^{\bd \ad}E^{\a\b\g}C_{\bd \ad}E_{\a\b\g}G(q^2) = E^{\a\b\g}E_{\a\b\g}G(q^2) \eea
 and, having extracted now sufficient momentum dependence, we need
only $G(0)$. (Since we are dealing with massive propagators $G$ is nonsingular at zero momentum).  
Furthermore, it is only necessary to calculate, diagram by diagram, contributions proportional to 
$q^2\d_{ab}$. It is then evident, looking at the structure of $A^\a$ and $B$ in 
(\ref{supergravity}) that we only need consider the term $e^{a \g}D_a\Del_\g$. The various 
possibilities are described then by the diagrams in Fig.3 with (a'), (b') omitted. We parametrize
the momentum dependence of the diagrams in the manner shown in Fig.4.  For example, Fig.3a leads to 
the following Feynman integral: \beq
 \int d^4p d^4k \frac{ (2p+q)_a (2p-q)_b}{ (p^2+m)^2
[(p+q)^2+m][(p+k)^2+m] (k^2+m)^3} \label{Feynman} \eeq

\vskip 18pt
\noindent
%---------- FIGURE TOP ------------
\begin{minipage}{\textwidth}
\begin{center}
\includegraphics[width=0.60\textwidth]{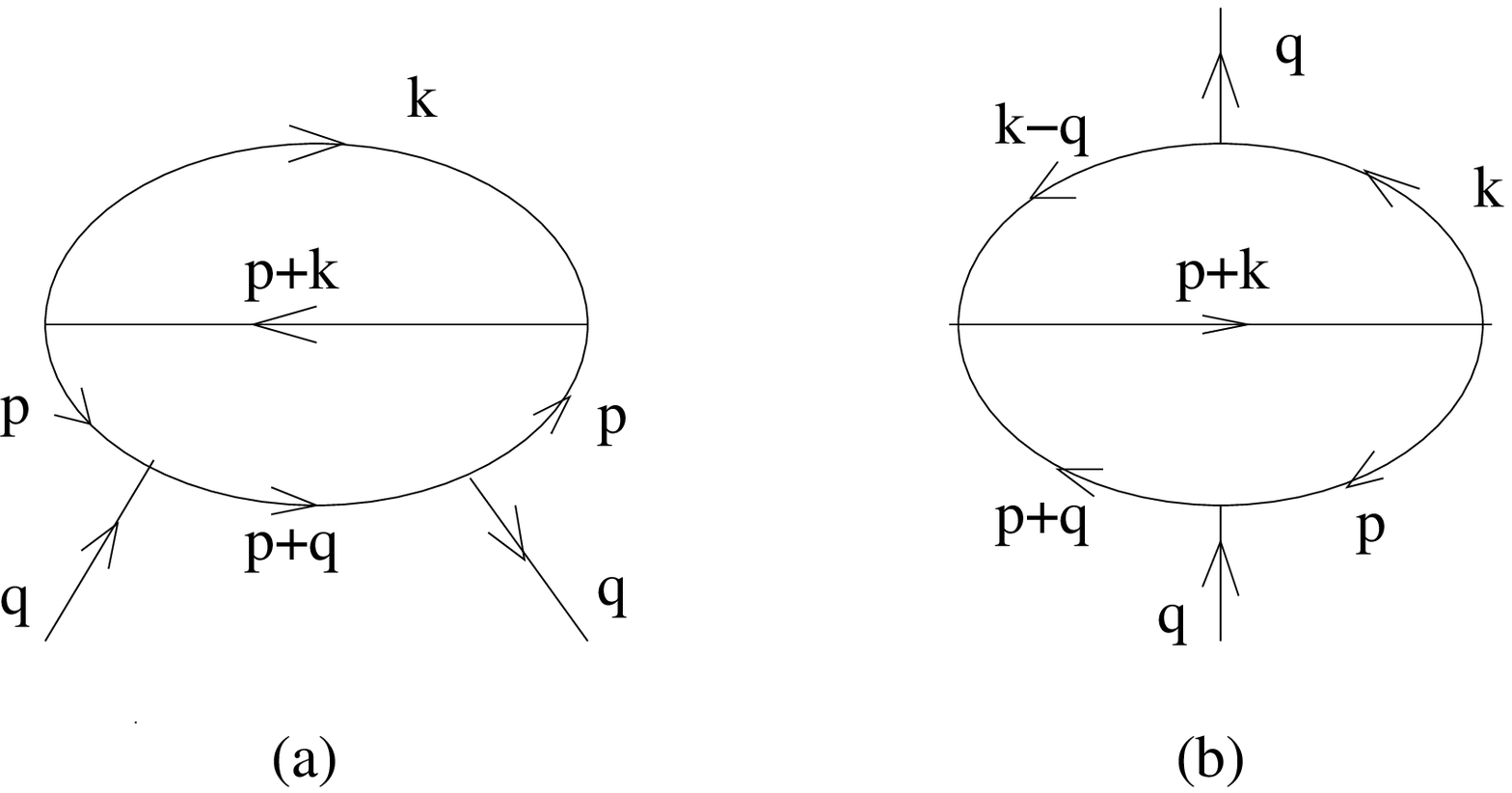}
\end{center}
\begin{center}
{\small{Figure 4: Momentum routing}}
\end{center}
\end{minipage}
%---------- FIGURE END ------------

\vskip 20pt

We can drop the $q$ factors since terms like $p_a q_b$ and $q_aq_b$ will never produce, after 
momentum integration, a result proportional to $\d_{ab}$. However, the momentum integration from 
$p_ap_b$ and $p_ak_b$ will give contributions to both $\d_{ab}$ and $q_aq_b$. To isolate the 
$\d_{ab}$ contribution we introduce therefore the operator \beq {\cal O}_{ab} = 2 
\frac{\partial^2}{\partial q^a \partial q^b} - 5 \frac{\partial^2}{\partial q^c
\partial q_c} \d_{ab}
\eeq 
which is such that ${\cal O}_{ab}q^aq^b=0$. Thus, if a particular
diagram produces, after momentum integration, an expression of the
form 
\beq 
I_{ab}^{(n)}=q^2
\d_{ab}G_1^{(n)}(q^2) +q_aq_b G_2^{(n)}(q^2)
\eeq 
we obtain {\it at zero momentum} 
\beq 
{\cal O}^{ab} I^{(n)}_{ab} |_{q^2=0} = -144 G_1^{(n)}(0) 
\eeq 
which is all that is needed, after summing over all the diagrams, to
obtain the desired result, $G(0)=\sum_1^6G_1^{(n)}$.

We proceed in the following manner: for each diagram we use a
Schwinger parameter representation of the propagators as follows:
\beq
{1\over \left(p^2+m\right)^{n+1}} = \int_0^{\infty} ds~\frac{s^n}{n!}
e^{-s(p^2+m)}
\eeq
Furthermore, we find
\bea
 {\cal O}_{ab} p^ap^b \left(
e^{-s(p+q)^2)}\right)_{q=0} &=& 12\left(3sp^2-s^2p^4\right)e^{-sp^2} 
\nonumber\\
{\cal O}_{ab} p^ak^b\left( e^{-s(p+q)^2}e^{-t(q-k)^2} \right)_{q=0}
&=&4\left[9(s+t)p\cdot k - 3s^2p^2k \cdot p-3t^2k^2p\cdot k\right.
\nonumber\\ &&\left.-2 st p^2k^2+8 st (p\cdot k)^2\right] e^{-sp^2-tk^2}
\eea
We note that, in general, we start with four or five Schwinger
parameters. For example, for the contribution in (\ref{Feynman}) we
must introduce separate factors
\beq
e^{-s_1(p^2+m)} ~~~,~~~~e^{-s_2[(p+q)^2+m]}
\eeq
for the propagators on the bottom line before applying the operator
${\cal O}_{ab}$ and then setting $q=0$. Afterwards, part of the
Schwinger parameter integration involves the integral
\beq
  \int ds_1 ds_2 s_1\left(3 s_2p^2 - s_2^2 p^4
  e^{-(s_1+s_2)(p^2+m)}\right)
\eeq
Changing variables to $s_1=xs ~,~ s_2=(1-x)s$ one can carry out the
integration over $x$ thus reducing the number of parameters. Other
diagrams must be dealt with in a similar manner.

To carry out the momentum integrations we write
\bea
 &&(i) ~~~  p.k e^{-sp^2-tk^2-u(p+k)^2} = -\frac{1}{2}  \left( \frac{\partial}{\partial
u}-\frac{\partial}{\partial
t}-\frac{\partial}{\partial s}\right) e^{-sp^2-tk^2-u(p+k)^2} \nonumber\\
&&(ii) ~~~ p^2 e^{-sp^2-tk^2-u(p+k)^2} =- \frac{\partial}{\partial s}
\left( e^{-sp^2-tk^2-u(p+k)^2}\right)\\
&&(iii) ~~~ p^4 e^{-sp^2-tk^2-u(p+k)^2}= \frac{\partial^2}{\partial s^2}
\left( e^{-sp^2-tk^2-u(p+k)^2}\right)\nonumber\\
&&(iv) ~~~ p^2 k^2e^{-sp^2-tk^2-u(p+k)^2}= \frac{\partial^2}{\partial s
\partial t } \left( e^{-sp^2-tk^2-u(p+k)^2}\right)\nonumber\\
&&(v) ~~~ p^2(p\cdot k) e^{-sp^2-tk^2-u(p+k)^2} = \frac{1}{2}\frac{\partial}{\partial s}\left(
\frac{\partial}{\partial u}
-\frac{\partial}{\partial s}-\frac{\partial}{\partial t}\right) \left( e^{-sp^2-tk^2-u(p+k)^2}\right)\nonumber\\
&&(vi) ~~~(p.k)^2 e^{-sp^2-tk^2-u(p+k)^2}= \frac{1}{4}\left( \frac{\partial}{\partial u}
-\frac{\partial}{\partial s}-\frac{\partial}{\partial t}\right)^2 \left(
e^{-sp^2-tk^2-u(p+k)^2}\right)\nonumber
 \eea
We can perform now the momentum integration
\beq
 \int d^4pd^4k\,
e^{-sp^2-tk^2-u(p+k)^2} = \left(\frac{1}{4 \pi}\right)^4 \frac{1}{[st +tu+us]^2} 
\eeq
after which we can carry out the differentiations with respect to the
Schwinger parameters leading to the following individual diagram
contributions:
\bea
&&Fig.(3a): ~~~ 4\frac{s^4t^4+2s^4t^3u+ s^4t^2u^2+2s^3t^3u^2+{\rm permutations}}{\Delta^4}
\nonumber\\
&&\nonumber\\
&&Fig.(3b): ~~~ \frac{2s^4t^3u+ 4s^4t^2u^2+8s^3t^3u^2+{\rm permutations}}{\Delta^4} \nonumber\\
&&\nonumber\\
&&Fig.(3c):~~~ -2\frac{-2s^4t^3u-6 s^4t^2u^2-4s^3t^3u^2+{\rm permutations}}{\Delta^4}\nonumber\\
&&\nonumber\\
&&Fig.(3d): ~~~-2 \frac{-s^4t^3u-2 s^4t^2u^2-10s^3t^3u^2+{\rm permutations}}{\Delta^4}\nonumber\\
&&\nonumber\\
&&Fig.(3e): ~~~ \frac{-s^4t^3u-2 s^4t^2u^2-\frac{10}{3}s^3t^3u^2+{\rm permutations}}{\Delta^4}
\nonumber\\
&&\\
&&Fig.(3f): ~~~ \frac{-s^4t^3u-2s^4t^2u^2-\frac{8}{3}s^3t^3u^2+{\rm permutations}}{\Delta^4}
\nonumber
\eea
all multiplied by the factor $e^{-m(s+t+u)}$.

In the above we have included the net factor from group theory,
D-algebra, and combinatorics that each diagram contribution must be
multiplied by. Additional overall factors are $N^2$, $(4\pi)^{-4}$
from the integration, and $- 1/144$ to take into account the factor
produced by the operator ${\cal O}_{ab}$.

Summing then all the contributions from Fig.3, the result takes the form \bea 
&&\frac{N^2}{144(4\pi)^4}~ E^{\a \b \g} E_{\a \b \g} {\rm Tr}{\cal W}^\d {\cal W}_\d \int ds dt du~ 
e^{-m(s+t+u)}
\nonumber\\
&&~~~~~~~~~\times \frac{4s^4t^4+14s^4t^3u+ 20s^4t^2u^2+38s^3t^3u^2+{\rm
permutations}}{\Delta^4}\nonumber\\
&&=\frac{N^2}{144(4\pi)^4}~ E^{\a \b \g} E_{\a \b \g} {\rm Tr}{\cal W}^\d {\cal W}_\d \int ds dt 
du~ e^{-m(s+t+u)}
\nonumber\\
&&~~~~~~~~~\times \frac{4s^2t^2+6s^2tu+{\rm permutations}}{\Delta^2} \label{EEWW}
\eea
We note that in the sum two factors of $\Delta$ have cancelled between
numerator and denominator.

Finally, we add together the contributions in (\ref{WWWW}) and
(\ref{EEWW}). Remarkably, just as in the pure SSYM case of \cite{DGLVZ}, in the sum
the denominator $\D^{-2}$ is
cancelled and after carrying out the now trivial integral over
Schwinger parameters the final result for this particular contribution
takes the form

\beq -\frac{N^2}{48(4\pi)^4 m^3}~  E^{\a \b \g} E_{\a \b \g} {\rm Tr}\,{\cal W}^\d {\cal W}_\d \eeq
a result consistent with that from string theory.

\newsection{Universality of the planar contribution}

{}From what we have seen the mixed glueball/gravitational F-terms from
genus zero takes the form
$$
{\cal L}= E^{\a \b \g} E_{\a \b \g} \sum_{i,j}
N_i N_j {\partial^2 {\cal F}_0 \over \partial S_i \partial S_j}
$$
where ${\cal F}_0$ is the planar partition function. We should note
that the full prepotential ${\cal F}_0$ also includes the measure
factor ${1\over 2} \sum_i S_i^2 \log S_i$ (which for the similar
expression for the superpotential yields the standard
Veneziano-Yankielowicz expression $\sum_i N_i S_i \log S_i$ for
the gauge factors $U(N_i)$). This measure factor should also be
included for the gravitational contribution, where it gives the term
proportional to
$$
\sum_i N_i^2 \log S_i.
$$
This is a direct consequence of the gravitational contribution to the
axial anomaly. %\cite{grava}
Note that the contribution of each gauge
factor $U(N_i)$ is proportional to $N_i^2$, since the gravitational
term in the anomaly keeps track of all the perturbative degrees of
freedom that are running around in the fermion loop. In this case
these are the $N_i^2$ components of the gluinos.

Now consider the expectation value of the gluino bilinear $S_i$ as
determined by extremization of the superpotential, which gives the
equation
\cite{pert}
$$N_i \partial_{i}\partial_{j}F_0+\tau=0,
$$
with $\tau$ the bare gauge coupling of the $U(N)$ gauge theory.

When the corresponding solutions for the $S_i$ are substituted into the gravitational
correction we have computed, one obtains
 therefore
$${\cal L}= -E^{\a \b \g} E_{\a \b \g} \sum_j N_j \tau.
$$
This  correction is proportional to the universal contribution $N\tau$ where
$N=\sum_jN_j$. It only depends on the total rank $N$ of the
gauge theory and is independent of the particular symmetry breaking
pattern and of all the details of the ${\cal N}=1$ superpotential. 

In fact, it is easy to see that this contribution, proportional to $N \tau$, is also needed for the
closed string dualities to work, if we embed these gauge theories into superstrings \cite{vaa,civ}:
For example consider Type IIB strings with some D5 branes wrapping 2-cycles of a CY and filling the 
spacetime.  Then for each brane there is a well-known $R\wedge R$ correction on its worldvolume 
\cite{bsv}.  Since the volume of the internal part of the $D5$ brane is given by $\tau$, this 
yields a term in four dimensions given by $\tau E^{\a \b \g} E_{\a \b \g}$.  Since we have $N$ such 
branes this gives exactly the contribution
$$E^{\a \b \g} E_{\a \b \g} N \tau.$$
In the context of superstrings this term comes in addition to the
glueball contributions we computed above. Thus, we now see that
the two effects -- the induced curvature term on the brane and the sum
of the planar diagrams of the gauge theory -- exactly cancel out. In
particular on the closed string dual, where the branes have
disappeared completely, there should be no genus zero correction to
$E^{\a \b \g} E_{\a \b \g}$; there should only be the genus one
contribution. This is indeed the case \cite{bcov,agnt}.

Note that, if we do not extremize the superpotential, the
gravitational correction receives contributions from both genus zero
and genus one diagrams. In cases when these diagrams can be exactly
summed and give rise to an effective spectral curve, as in
\cite{dvi,dvii}, these contributions have a direct geometric
interpretation. The genus zero quantity $\partial_i\partial_j{\cal
F}_0$ gives the period matrix $\tau^{\rm eff}_{ij}$ of the effective
curve \cite{pert}, and the genus one term ${\cal F}_1$ can be
expressed as  the chiral scalar determinant \cite{dijk,mark}. Combining
these two facts, we can write the gravitational correction as $\log Z$
with
$$
Z = {e^{\pi i N_i \tau^{\rm eff}_{ij} N_j} \over \sqrt{\det \Delta}}.
$$
We note the amusing fact that $Z$ takes the form of a holomorphic
block of a chiral boson on the spectral curve with loop momenta $N_i$.

\subsection{Gravitational F-term as domain wall
partition function}

As we have argued the whole non-trivial contribution of the
gravitational F-terms will come from genus one diagrams, {\it i.e.}
$$
{\cal L}= {\cal F}_1(S_i) E^{\a \b \g} E_{\a \b \g}$$
where we substitute the value of $S_i$ found from the extremization of
the superpotential, as computed using the planar diagrams. It is
natural to ask what is the ${\cal N}=1$ supersymmetric gauge theory
significance of this term. Let us think of this as if it were to come
from a dual closed string theory. In this context we see that $N$
does not enter this expression, so it would have made sense also for
${\cal N}=2$ theories, where the flux, which breaks half of the
supersymmetries and  is proportional to $N$, is set to zero. In
the context of ${\cal N}=2$ supersymmetric theories obtained by type
IIB strings on CY 3-folds, it has been argued in \cite{volt}\ that the
genus one term ${\cal F}_1$ computes the partition function of BPS D3
branes wrapped over cycles of the Calabi-Yau. Roughly speaking we have
$$
{\cal F}_1={1\over 12}\sum_{\hbox{\small \it BPS states}}
 (-1)^s \log m,
$$
where $s$ denotes the spin of the D3 brane state and $m$ denotes its
mass, given by $|\int_{D3} \Omega|$. This is only roughly the
description because the number of D3 branes can jump over moduli
whereas ${\cal F}_1$ is smooth. This is because ${\cal F}_1$ includes
also contributions from multi-particle sectors of D3 branes as in
\cite{cfiv}. It would be interesting to make this interpretation of the gravitational correction as counting BPS states more precise.

However, for the case at hand, with generically just ${\cal N}=1$ supersymmetry, this is not a 
satisfactory interpretation of the gravitational correction, because there is no notion of BPS 
particle for ${\cal N}=1$ supersymmetric gauge theories.  The only BPS object is the domain wall. 
In the string setup this is related to D5 branes which wrap a 3-cycle inside a CY and are a domain 
wall in ${\bf R}^4$.  Since the internal part of the counting of these BPS domain walls is the same 
as the counting of BPS particles in the associated ${\cal N}=2$ supersymmetric theory, it is 
natural to conjecture that the ${\cal N}=1$ supersymmetric interpretation of gravitational F-term 
is as a partition function of domain walls.

\bigskip
\bigskip

\centerline{{\bf Acknowledgments }}

\bigskip

C.V. thanks the hospitality of the theory group at Caltech, where he
was a Gordon Moore Distinguished Scholar. R.D., H.O., and C.V. thank
the Simons Workshop on Mathematics and Physics and the YITP at Stony
Brook for their hospitality during the completion of this work.

The research of R.D.~was partly supported by FOM and the CMPA grant of the University of Amsterdam. 
The research of M.T.G. was supported by NSF grant PHY-0070475 and NSERC grant 204540.
The research of H.O. was supported in part by DOE grant DE-FG03-92-ER40701.  The research of C.V.
was supported in part by NSF grants PHY-9802709 and DMS-0074329.  The research of D.Z. was 
supported in part by INFN, MURST, and the European Commission RTN program HPRN-CT-2000-00113 in 
which the author is associated to the University of Torino.

\renewcommand{\Large}{\normalsize}

\end{document}